\documentclass[twocolumn]{aastex631}
\usepackage{bm}
\usepackage{amsmath}
\graphicspath{./}

\begin{document}

\title{Black hole mergers driven by a captured low-mass companion }

\author[0000-0003-2270-1310]{Stephen Lepp}

\author[0000-0003-2401-7168]{Rebecca G. Martin}

\author[0000-0002-9725-2524]{Bing Zhang}

\affiliation{Nevada Center for Astrophysics, University of Nevada, Las Vegas, 4505 S. Maryland Pkwy., Las Vegas, NV 89154, USA}
\affiliation{Department of Physics and Astronomy,University of Nevada, Las Vegas, 4505 S. Maryland Pkwy., Las Vegas, NV 89154, USA}

\begin{abstract}
Increased eccentricity of a black hole binary leads to reduced merger times. With $n$-body simulations and analytic approximations including the effects of general relativity (GR), we show that even a low mass companion orbiting a black hole binary can cause significant eccentricity oscillations of the binary as a result of the Kozai-Lidov mechanism.   
A companion with a mass as low as about $1\%$ of the binary mass can drive the binary eccentricity up to $\gtrsim 0.8$, while a mass of a few percent can drive eccentricities greater than 0.98. For low mass companions, this mechanism requires the companion to be on an orbit that is closer to retrograde than to prograde to the binary orbit and this may occur through capture of the third body.    The effects of GR limit the radial range for the companion for which this mechanism works for the closest binaries.  The merger timescale may be reduced by several orders of magnitude for a captured companion mass of only a few percent of the binary mass.
\end{abstract}


\section{Introduction} \label{sec:intro}

Observations of gravitational-wave emission from the mergers of black holes \citep[e.g.][]{Abbott2016} are challenging our understanding of the formation of black hole binaries. The initial separation required for the binary to merge through gravitational wave emission in less than the age of the Universe is only about $0.23\,\rm au$ for an equal mass black hole binary with a total mass of $65\,\rm M_\odot$ \citep[e.g.][]{Silsbee2017}. Formation of such close binaries may occur through a common envelope phase during which the orbit shrinks \citep[e.g.][]{Belczynski2002}, dynamical formation in globular clusters \citep[e.g.][]{Rodriguez2016}, or formation in the gas-rich AGN disk \citep[e.g.][]{Bartos2017,Li2023}. 

High eccentricity of black hole binaries can speed up their merger rate and add additional signatures to the gravitational waves \citep[e.g.][]{xuan2022}.  Binary eccentricity growth may be driven by
Kozai-Lidov (KL) oscillations \citep{Zeipel1910,kozai1962,lidov1962} that are induced either by a nearby supermassive black hole \citep{antonini2012,Hoang2018} or intermediate mass black hole tertiary \citep{VanLandingham2016}, or driven in   an isolated black hole triple \citep{Silsbee2017}. When the orbit of the black hole binary is highly misaligned to its orbit around the supermassive black hole, or the tertiary orbit, the inner black hole binary exchanges angular momentum with the outer companion, leading to opposing oscillations of the inner orbit eccentricity and inclination \citep[e.g.][]{Naoz2016,Hamers2021}. 

Many authors have studied black hole mergers by KL oscillations 
\citep{VanLandingham2016,kimpson2016}
and recent calculations have explored the direct detection of such oscillations in merging black holes \citep{hoang2019,barnabas2020}.
For wide binaries (greater then $1\,\rm au$) with  a massive third body, with a mass $\gtrsim 0.1$ times the binary mass, the eccentricity growth is maximal when the third body is in a polar orbit around the inner binary \citep{Liu2018,Liu2019,Liu2019b}.
 In this work we explore the feasibility of driving black hole mergers with a low-mass, but close-by, companion object. 

The formation of a black hole binary with a tertiary companion could occur through similar processes to the formation of the binary itself. The tertiary could be captured in a gas-rich environment of a common envelope or an AGN disc, or the capture could occur dynamically.
The capture cross section for two black holes in an elliptical orbit scales as the mass of the scattering object \citep{Valtonen2006} and so the cross section for capturing a $30\,\rm M_\odot$ object would be about 30 times that for capturing a $1\,\rm M_\odot$. However, the relatively low mass objects may be much more numerous. If the objects available  scale with the Salpeter initial mass function \citep{salpeter1955} then $1\,\rm M_\odot$ captures occur at three times the rate of $30\,\rm M_\odot$ captures.

There are two types of stationary circumbinary particle orbits (meaning orbits that do not undergo nodal precession) around eccentric orbit binaries:  coplanar and polar to the binary orbit \citep{Verrier2009,Farago2010,Doolin2011,Naoz2017,Martin2019,Chen2019}. In a polar orbit, the particle orbits perpendicular to the binary orbital plane, aligned to the binary eccentricity vector. Aside from the stationary orbits, particle orbits undergo nodal precession.  For low initial inclinations, particle orbits are in {\it circulating} orbits, meaning that the particle angular momentum vector precesses about the binary angular momentum vector. For higher initial inclinations, particle orbits are {\it librating}, meaning that the binary angular momentum vector precesses about the polar stationary inclination. 

Without the effects of GR, the type of orbit does not depend upon the orbital radius of the particle. However, GR causes apsidal precession of the binary and the behaviour changes with the separation of the particle \citep[e.g][]{Zanardi2018}. For larger orbital radii, the stationary polar inclination increases from $90^\circ$  and the librating orbits move to higher inclinations. Therefore there is a maximum orbital radius for which the particle can be in a librating orbit \citep{lepp2022}.

In this Letter we investigate the range of possible parameters for which a low mass third body can drive a merger.  Even a relatively low mass third body can lead to large eccentricity oscillations of the binary \citep{Naoz2017,Chen2019,Abod2022}.
In Section~\ref{sec:three}, we first consider the dynamics of the triple system including the effects of GR. The outer body has a relatively small mass compared to the binary.  The largest eccentricity is driven from the librating orbit that begins farthest from the stationary inclination, closest to retrograde. 
We then constrain the parameters of the triple systems that can drive a merger. For a low mass third body, we find that the maximal eccentricity growth occurs for a retrograde orbit of the third body. In Section~\ref{analytic}, we compare our numerical simulations to an analytic model. In Section~\ref{merger} we consider the merger speed up and we draw our conclusions in Section~\ref{concs}.

\section{Three-body Simulations} \label{sec:three}

We use the REBOUND n-body code along with gr\_full package from REBOUNDx \citep{rebound, reboundx} to model a binary black hole with a misaligned low-mass companion.  The simulations were integrated using WHFast, a symplectic Wisdom-Holman integrator \citep{reboundwhfast,wh}, some  simulations were checked by integrating with the IAS15, a 15th order Gauss-Radau integrator \citep{reboundias15}. The SimulationArchive format was used to store fully reproducible simulation data \citep{reboundsa}.

For our standard model we consider a binary of two 
black holes with mass $m_1=m_2=30 \,\rm M_\odot$ for a total binary mass of $m_{\rm b}= 60\,\rm M_\odot$. These parameters are motivated by black hole merger observations.The first detected  black hole merger event, GW150914 \citep{Abbott2016}, involved two 
black holes each of roughly $30 \,\rm M_\odot$ and it was only in the third 
LIGO run that the twelfth black hole merger event, GW190412, had a mass ratio inconsistent
with equal mass \citep{abbott2020} with the merger of a $30 \,\rm M_\odot$ and an $8 \,\rm M_\odot$ black hole.  We also explore this case.

The black holes are in an orbit with a semi-major axis of $a_{\rm b}$ and 
an initial eccentricity of $e_{\rm b}=0.2$. The orbital period is $P_{\rm b}$.

For the semi-major axis, we consider a range of values, $a_{\rm b}=10$, $30$, $50$, and  $250\,\rm  R_\odot$.  In isolation, these systems have merger times of $0.8$, $70$, $500$,  and $300,000\,\rm  Gyrs$, respectively which bracket the Hubble time. The merger time calculation is discussed in section~\ref{merger}. 
We have chosen a small eccentricity of $e_{\rm b}=0.2$ to start, but find our results 
are relatively insensitive to this initial eccentricity.We discuss in Section~\ref{scaling} how our results scale to different binary masses and semi-major axes.

Around the binary  is a companion
orbiting in an initially circular orbit with an orbital radius which we vary 
from $r=5$  to $20\,a_{\rm b}$ and whose mass we vary from 
$m=0.5$ to  $4\,\rm M_\odot$. 
For our standard parameters we take the eccentricity of the outer body to be zero, however, we also consider the effect of the eccentricity of the outer body up to $e=0.8$.
The inclination of the companion orbit relative to that of 
the black hole binary is
\begin{equation}
i = \cos^{-1}(\hat{\bm{l}}_{\rm b}\cdot \hat{\bm{l}}_{\rm c})\,,
\end{equation}
where $\hat{\bm{ l}}_{\rm b}$ is a unit vector 
in the direction of the black hole binary angular momentum and 
$\hat{\bm{ l}}_{\rm c}$ is a unit vector in the direction of the  
angular momentum of the companions orbit. 
The nodal  phase angle of the companion is the angle measured relative to 
the eccentricity vector of the black hole binary and is 
given by 
\begin{equation}
        \phi = \tan^{-1}\left(\frac{\hat{\bm{l}}_{\rm c}\cdot (\hat{\bm{l}}_{\rm b}\times 
    \hat{\bm{e}}_{\rm b})}{\hat{\bm{l}}_{\rm c}\cdot \hat{\bm{e}}_{\rm b}}\right) + 90^\circ 
\end{equation}
\citep{Chen2019,Chen2020e}, where  $\hat{\bm{e}}_{\rm b}$ is the  unit eccentricity vector of the black hole  binary.

\subsection{Binary eccentricity evolution}

\begin{figure}
    \centering
    \includegraphics[width=\columnwidth]{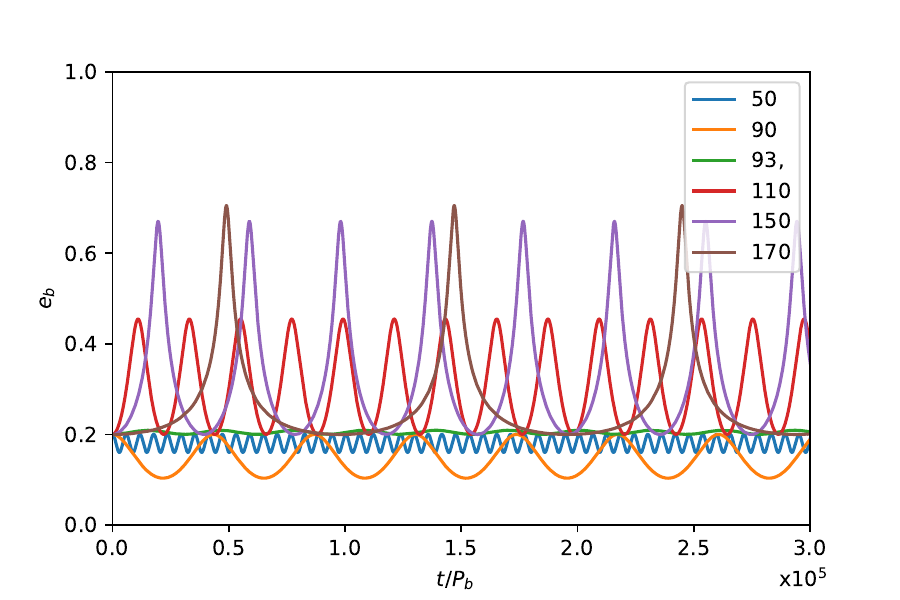}
    \caption{Eccentricity evolution of a black hole binary with masses $m_1=m_2=30\,\rm M_\odot$ orbiting with semi-major axis $a_{\rm b}=10\,\rm R_\odot$ and initial eccentricity $e_{\rm b}=0.2$. The third body has mass $m=1\,\rm M_\odot$ and orbits with semi-major axis $r=10\,a_{\rm b}$ with initial inclination of $i=50$, $90$, $93$, $110$, $150$ and $170^\circ$ and initial nodal phase angle $\phi=90^\circ$.}
    \label{fig:ecc}
\end{figure}

We begin by considering the behavior of the black hole binary eccentricity in response to a companion of mass $m=1\,\rm M_\odot$ orbiting at a radius of
$r=10\, a_{\rm b}$  including the effects of GR.  The outer companion  has a mass of only $1.7\%$ of the binary but an angular momentum of about $8\%$ of the inner black hole binary.  Fig.~\ref{fig:ecc} shows that this is large
enough to induce significant changes in the black hole binary eccentricity as a result of KL oscillations.
 The magnitude  and the initial direction of the oscillations depends upon the initial inclination. 
For initial inclinations less than the stationary inclination, $0^\circ<i<93^\circ$, the binary eccentricity  oscillates to lower eccentricity.
An inclination of $i=93^\circ$ is close to the stationary inclination and here 
the companion stays at this inclination and the black hole binary 
eccentricity remains roughly constant.
For initial inclinations larger than the stationary inclination, $i> 93^\circ$, the binary oscillates to higher  eccentricity, reaching a maximum of about $0.7$ for an initial inclination of $i=170^\circ$.

\subsection{Dynamics of the three body systems}

\begin{figure*}
\includegraphics[width=1.2\columnwidth]{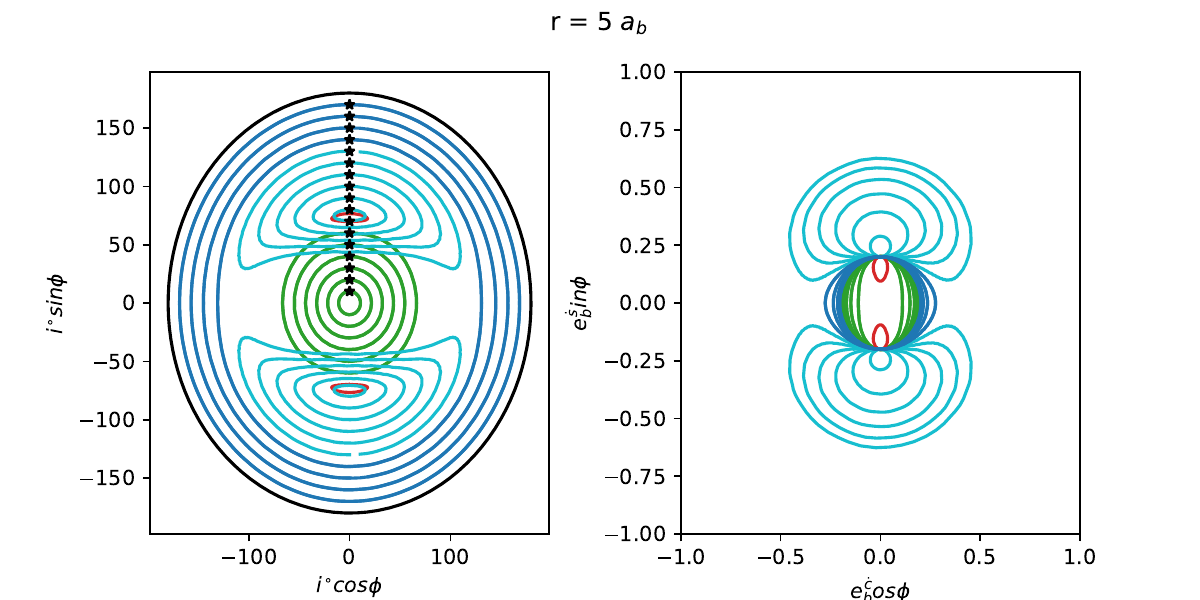}\hspace{-0.5in}
\includegraphics[width=1.2\columnwidth]{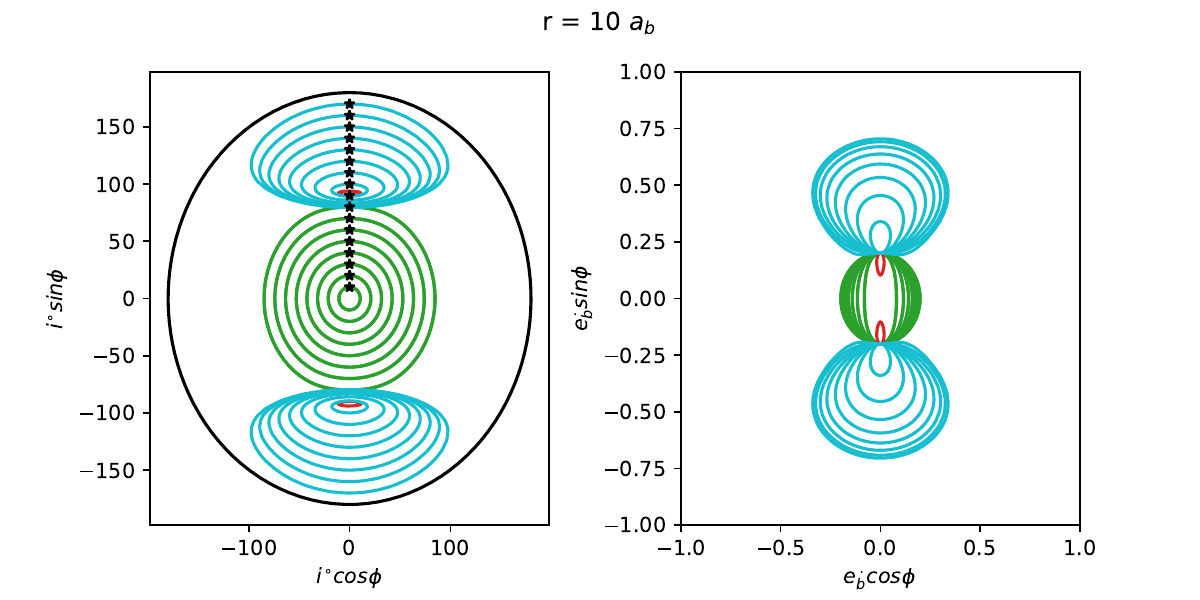}
\includegraphics[width=1.2\columnwidth]{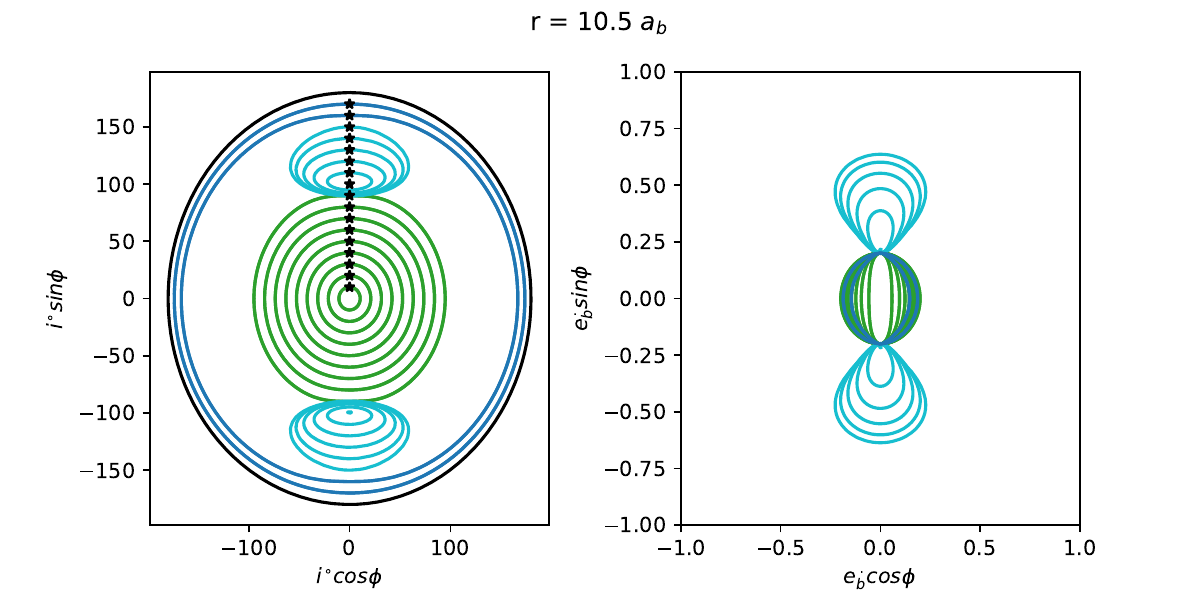}\hspace{-0.5in}
\includegraphics[width=1.2\columnwidth]{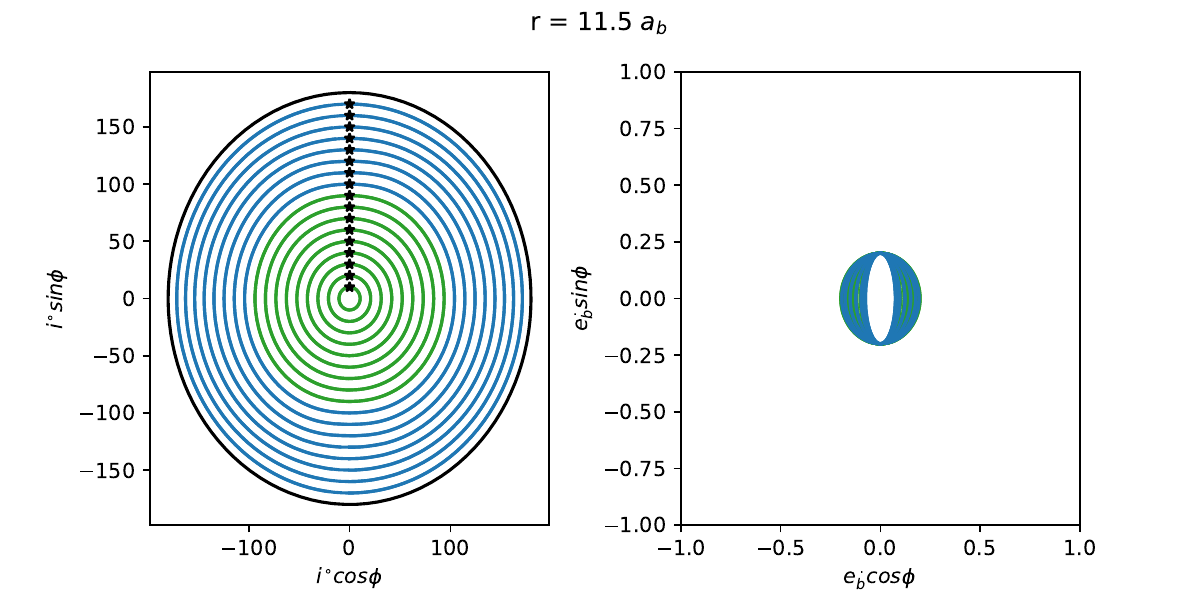}
\caption{The $i\cos \phi-i\sin \phi$ phase plane (first and third columns) and the $e_{\rm b}\cos \phi-e_{\rm b}\sin \phi$ phase plane (second and fourth columns) for a binary composed of $m_1=m_2=30\,\rm M_\odot$ with orbital separation of $a_{\rm b}=10\,\rm R_\odot$ and initial binary eccentricity $e_{\rm b}=0.2$. The companion has mass $m=1\,\rm M_\odot$ and orbits the binary in an initially circular orbit with orbital radius of radii $r=5$ (top left), $10$ (top right), $10.5$ (lower left) and $11.5 \,a_{\rm b}$ (lower right).  The third body begins with $\phi=90^\circ$ with inclination in the range $0-180^\circ$ with an interval of $10^\circ$. The green lines show prograde circulating orbits and the blue lines show retrograde circulating orbits. The red and cyan lines show librating orbits with initial inclination below and above the stationary inclination, respectively. The black stars show the initial conditions.
}
\label{fig:rs20}
\end{figure*}

Fig.~\ref{fig:rs20} shows $i \cos \phi-  i \sin \phi$ and 
$e_{\rm b} \cos\phi - e_{\rm b}\sin\phi$ phase diagrams for an equal mass black hole binary with  $m_1=m_2=30\,\rm M_\odot$  with semi-major axis $a_{\rm b} = 10\,\rm  R_\odot$ and an initial 
eccentricity of $e_{\rm b}=0.2$.  The companion mass is
$m=1\,\rm M_\odot$ and its orbital radius is  
$r=5$, $10$, $10.5$ and $11.5\,a_{\rm b}$.  
These plots show the evolution of the orbital elements of the 
companion relative to the black hole binary. 

The orbits are either circulating (nodally precessing about the binary angular momentum vector) or librating (nodally precessing about a non-zero stationary inclination) depending on their initial inclination.
The initial conditions are shown with a black star. Only orbits originating in the upper half of the phase diagram were calculated and the phase diagram was reflected to produce the lower half.
The upper left pair of panels  have a companion  at $r=5\,a_{\rm b}$.  In the inclination phase panel, for increasing initial inclination, starting at $i=10^\circ$, there are 6 circulating 
orbits (green) followed by 7 librating orbits. The first librating orbit has an initial inclination less than the stationary inclination (red) and the 
others have initial inclination greater than the stationary inclination (cyan). At even higher initial inclination there are four retrograde 
circulating orbits (blue).  
The $e_{\rm b} \cos\phi - e_{\rm b}\sin\phi$ plane shows how the binary eccentricity changes. For orbits with  an initial 
inclination smaller than the stationary 
incliation, the binary eccentricity oscillates to lower eccentricity while 
those with higher initial inclination oscillate to higher eccentricity.  For a close in companion, the effects of GR are not important.

The upper right pair of panels shows
the phase diagrams for a companion that has a larger orbital radius of
$r=10\,a_{\rm b}$. Here, the effects of GR become important. Without GR, this would be similar to 
the $r=5 a_{\rm b}$ panels.  Instead, the retrograde circulating regions 
have entirely disappeared and we are left with a large librating region which can drive the inner black hole binary to large eccentricity. The maximum eccentricity growth occurs for the librating orbit that begins with the largest inclination. For the orbits shown in the figure, this is $i=170^\circ$.

The panels for $r=10.5\, a_{\rm b}$ and $r=11.5 \,a_{\rm b}$ are shown in the second row.
The librating
region, which causes the large changes in black hole eccentricity,
has shrunk in the $10.5 \,a_{\rm b}$ panels and 
is not present for    $r=11.5\,a_{\rm b}$ panels.
This is similar to that seen in \cite{lepp2022}, except the 
transition occurs much more rapidly with radius for a third body with mass. 
This is because the binary apsidal precession from GR is modified by the apsidal precession 
caused by the massive companion \citep[e.g.][]{Lepp2023}. GR always causes the inner binary to precess in a prograde direction, while the direction of the precession induced by companion depends upon its inclination.  At inclinations near to $90^\circ$ the
precession caused by the companion is retrograde \citep[e.g.][]{Zhang2019,Childs2023}, whereas close to $i=0$ and $i=180^\circ$ it is prograde.   The stationary inclination  increases rapidly for companions with a mass of about $1\,\rm M_\odot$ and a separation of about $10\, a_{\rm b}$, because as the companion approaches $i=180^\circ$, the precession of the binary speeds up.  For larger mass 
companions the dynamics can be even more complicated as there may be multiple 
stationary states \citep[e.g.][]{Zanazzi2018,Chen2019,Abod2022}.

The time-scale for  KL oscillations to increase the eccentricity is very rapid compared with the merger time of the black hole.  For our standard case of two $30\,\rm M_\odot$ black holes with an $a_{\rm b}=10\,\rm  R_\odot$ and a $1\,\rm M_\odot$ companion orbiting at a $r=10 \,a_{\rm b}$ the maximum eccentricity of the black hole binary occurs for the $i=170^\circ$ initial inclination of the companion.  It reaches this eccentricity in 65 years.

\subsection{Maximum eccentricity of the binary}

\begin{figure*}
\begin{center}
\includegraphics[width=0.9\columnwidth]{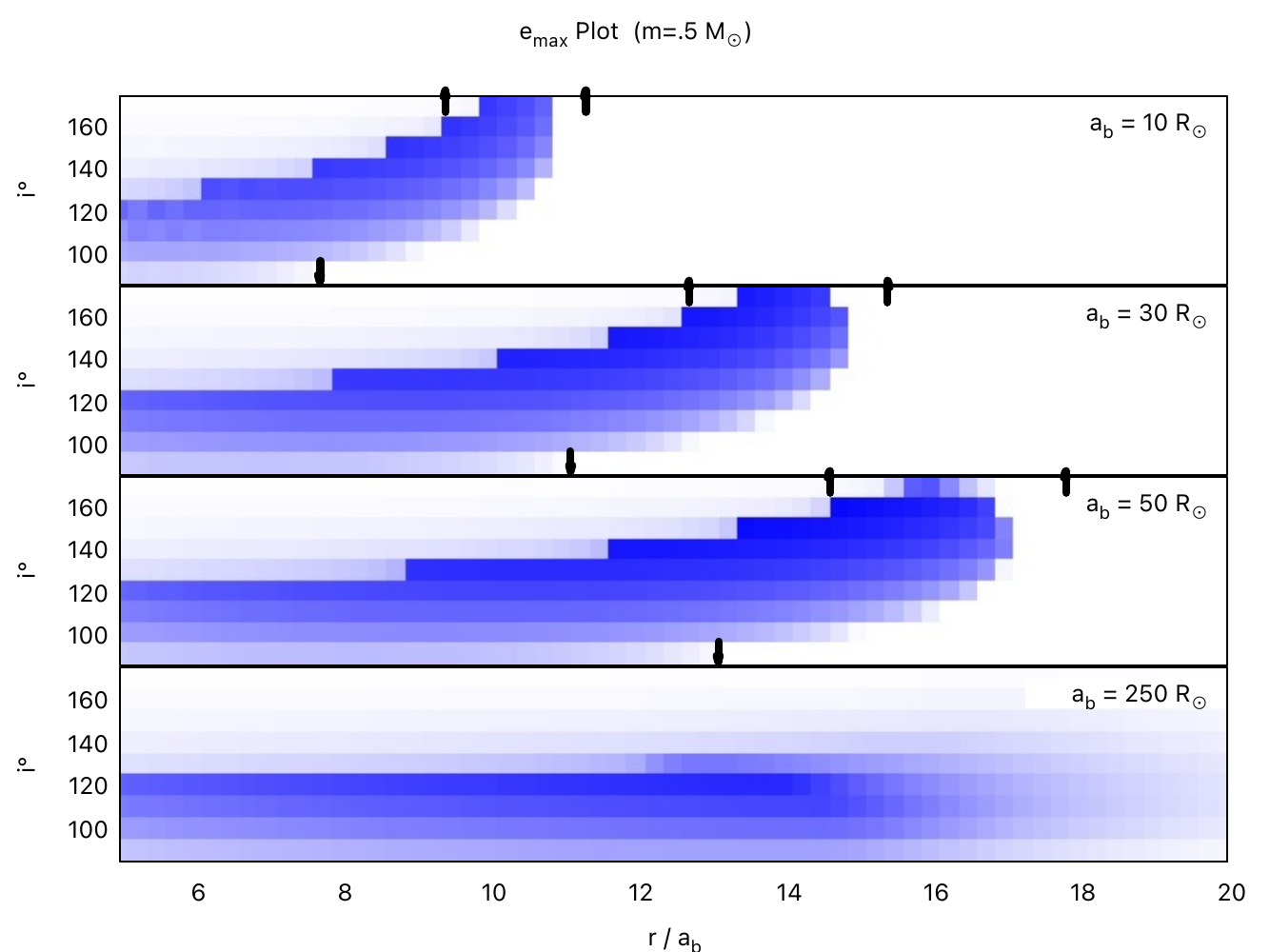}
\includegraphics[width=0.9\columnwidth]{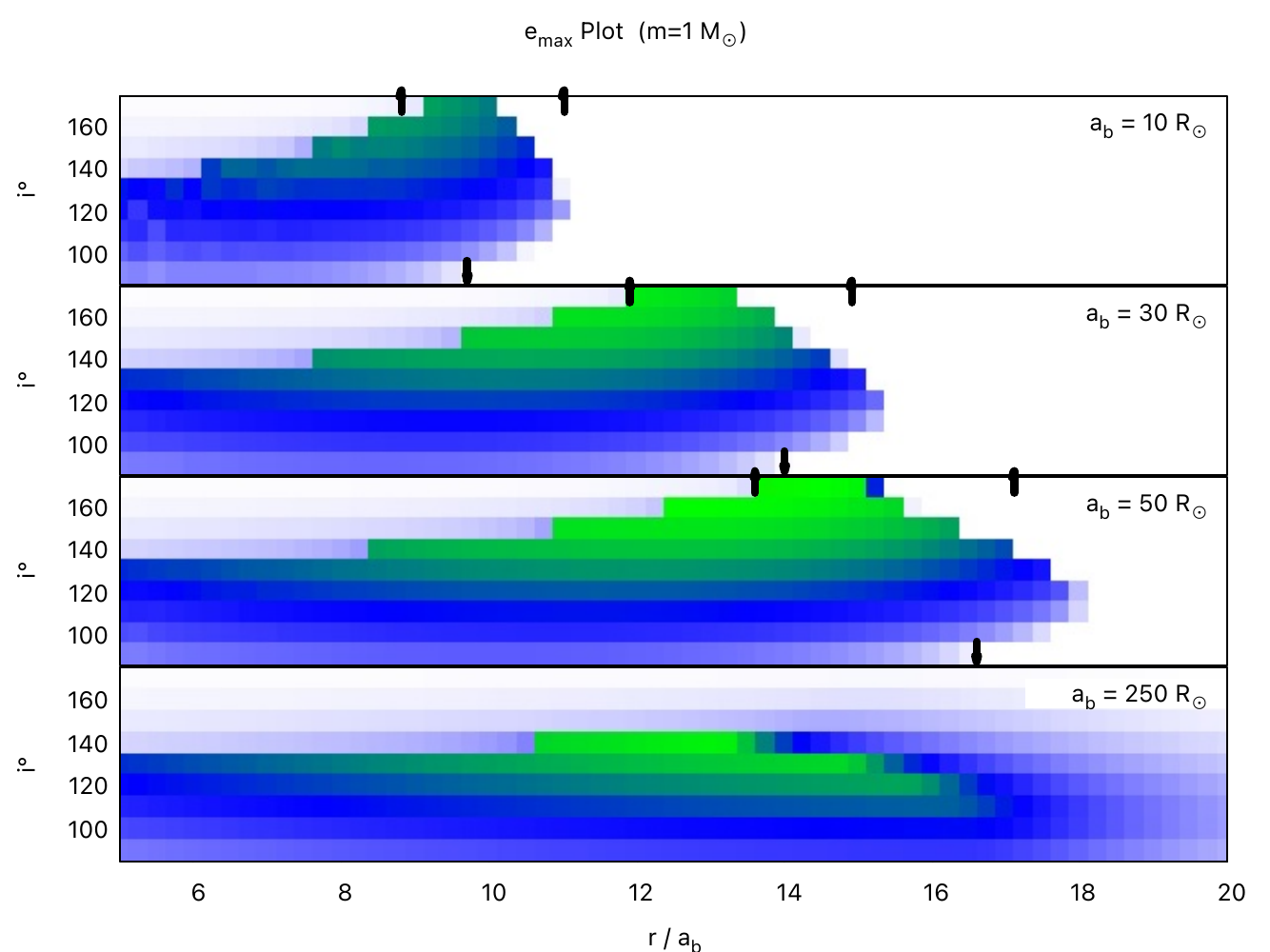}
\includegraphics[height=.6\columnwidth]{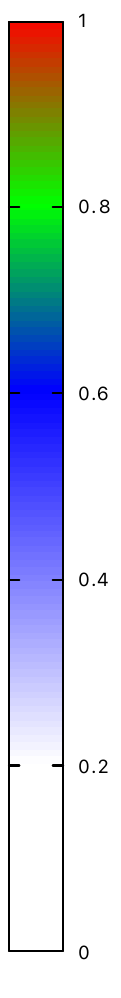}\raisebox{1in}{$e_{\rm max}$}
\includegraphics[width=0.9\columnwidth]{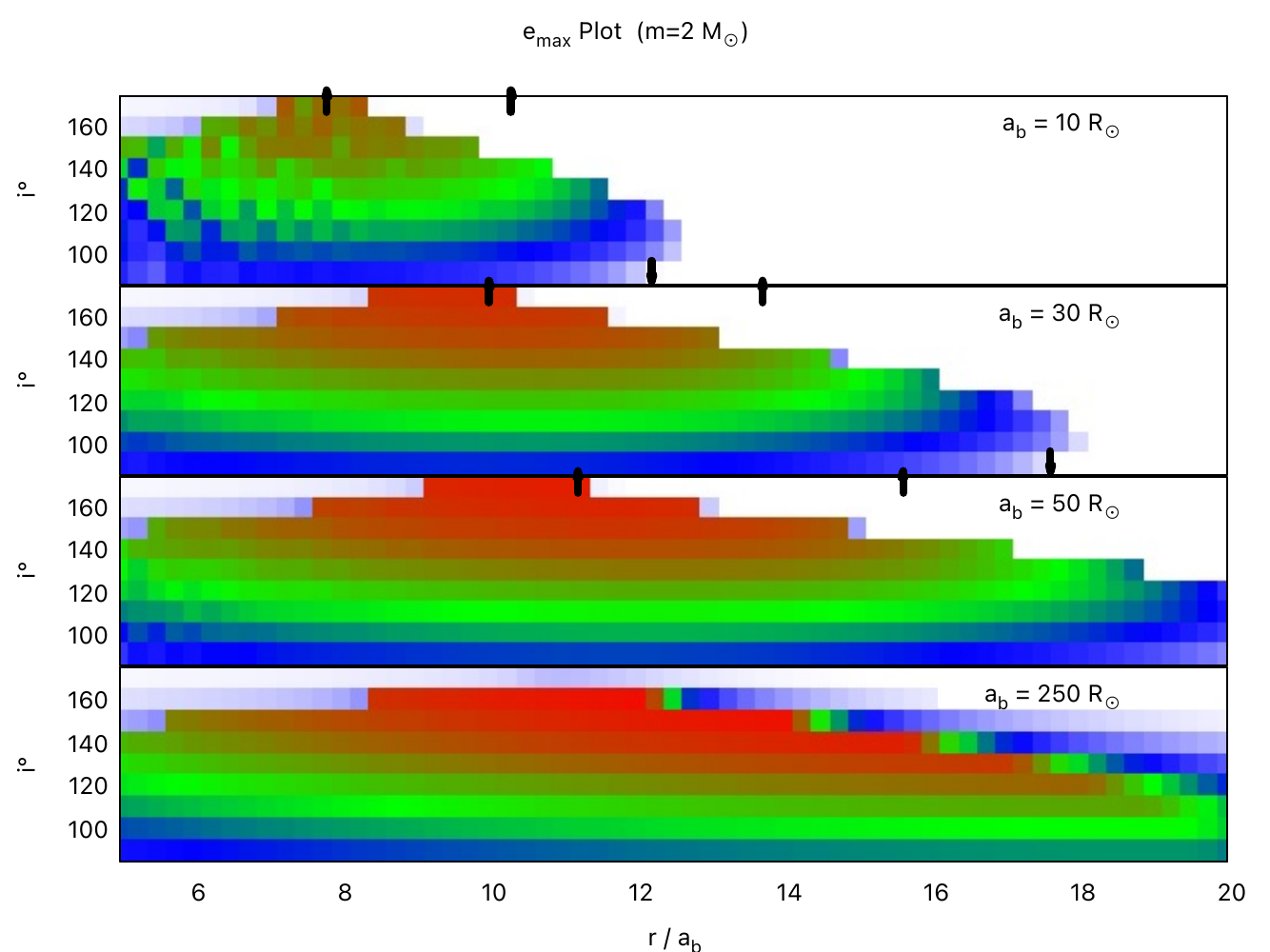}
\includegraphics[width=0.9\columnwidth]{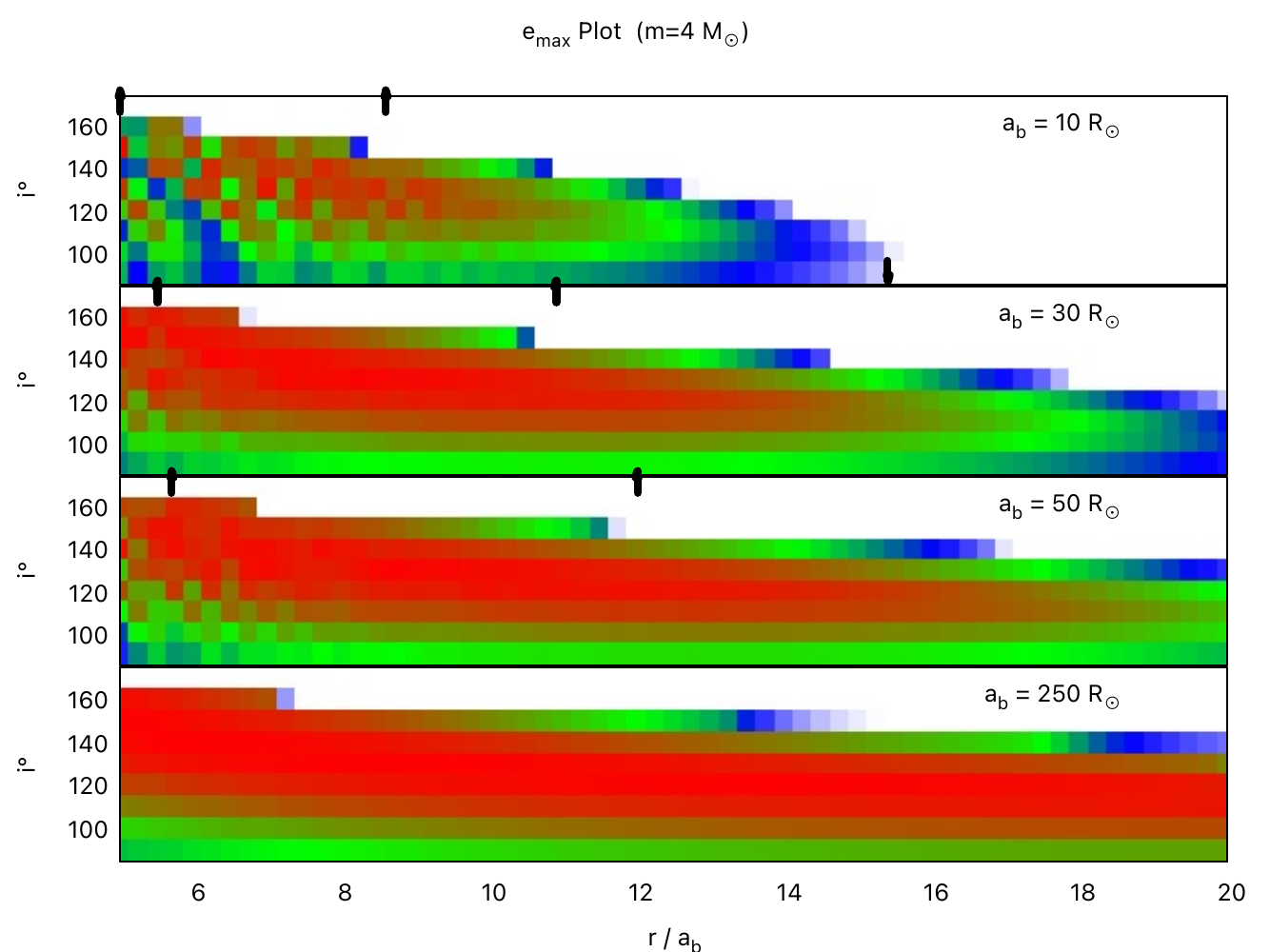}
\includegraphics[height=.6\columnwidth]{color.pdf}\raisebox{1in}{$e_{\rm max}$}
\includegraphics[width=0.9\columnwidth]{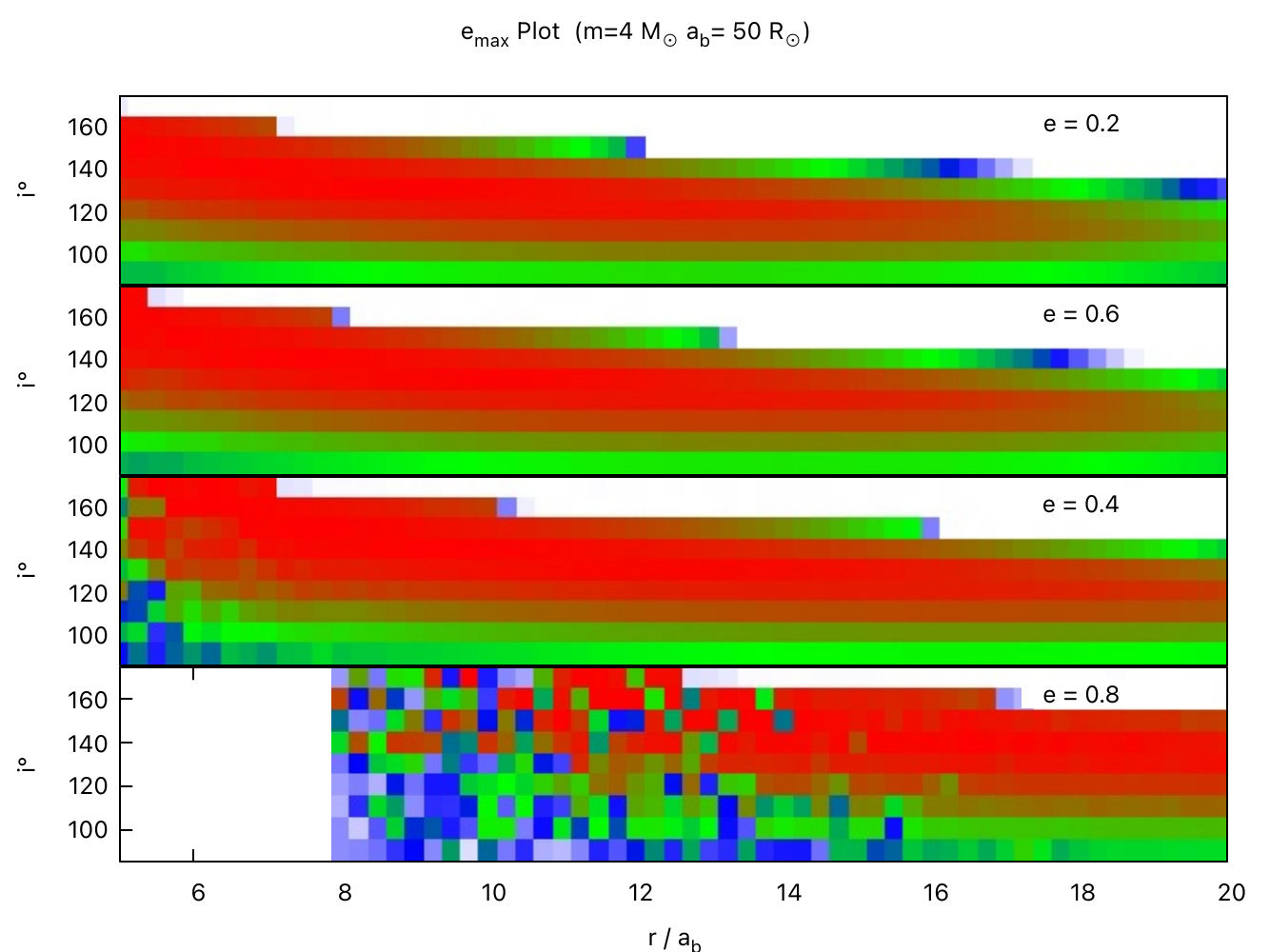}
\includegraphics[width=0.9\columnwidth]{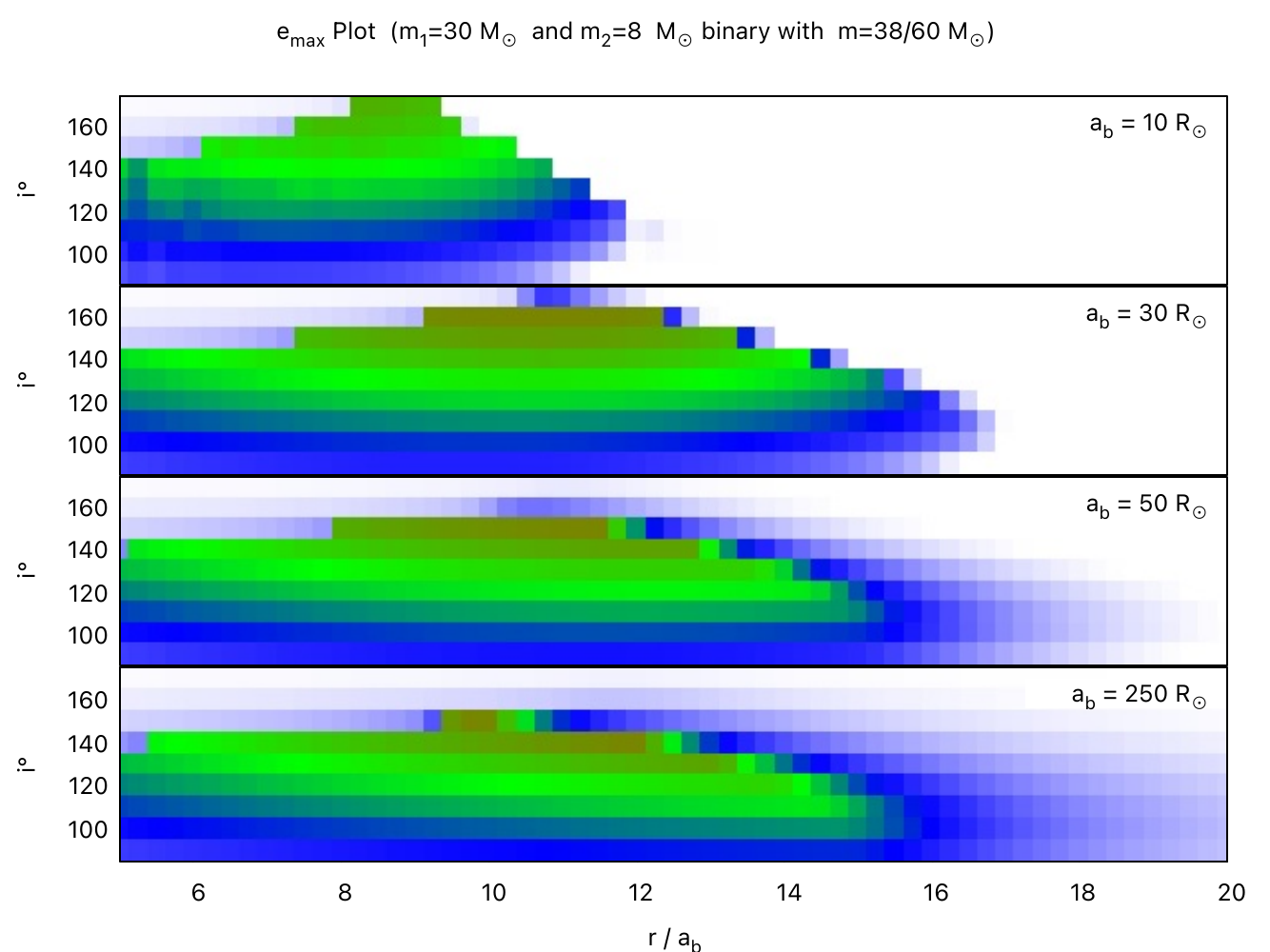}
\includegraphics[height=.58\columnwidth]{color.pdf}\raisebox{1in}{$e_{\rm max}$}
\end{center}
\caption{Maximum eccentricity, $e_{\rm max}$, for varying initial inclination, $i$, and orbital radius of the companion, $r$. The upper four panels each show
 $a_{\rm b}=10$, $30$, $50$ and $250\,\rm R_\odot$ and for a  companion mass $m=0.5$ (top left), $1$ (top right), $2$ (middle left) and $4\,\rm M_\odot$ (middle right).  The lower left panel explores the effect of the eccentricity of the tertiary mass with $a_{\rm b}=50\,\rm R_\odot$ and $m=4\,\rm M_\odot$. The lower right panel is the same as the upper right panel except $m_1=30\,\rm M_\odot,$ $m_2=8\,\rm M_\odot$ and $m=38/60 \,\rm M_\odot$.  The radii where the librating region ($r_{\rm c2}$, left arrow) and stationary inclination ($r_{\rm c1}$, right arrow) reach $i=180^\circ$ are marked by black arrows at $i=180^\circ$. The location where the GR precession is just canceled by the three body precession for the blackhole binary (where $i_{\rm s}=90^\circ$), $r_{\rm c3}$, is marked by a black arrow at $i=90^\circ$.  Note that for $a_{\rm b}=250\,\rm R_\odot$ all the arrows are off the right side of the plot.  }
\label{fig:contour}
\end{figure*}

Fig.~\ref{fig:contour} shows the maximum eccentricity of the binary that is reached as a function of the initial inclination and orbital radius of the third body.  A grid of models was run for four different tertiary masses $m= 0.5$, $1$, $2$, and  $4\,\rm M_\odot$ (upper four panels).  Each mass was run at different initial inclinations from 
$90$  to  $170^\circ$ in steps of $10^\circ$ and for different
radii from $5$ to  $20\, a_{\rm b}$ in steps of $0.25\, a_{\rm b}$.   
The top left panel shows the results for a companion mass 
of $m=0.5\,\rm M_\odot$.  There is a significant region where
the maximum eccentricity reaches about 0.6.  
The 
$m=1\,\rm M_\odot$ plot (top right) is very similar although for some parameters the maximum eccentricity has increased to about $0.8$. 
For larger companion mass (middle panels), there are regions where the maximum value
of $e_{\rm b}$ is even larger, up to almost $e_{\rm b}=1$.

There are two main mechanisms driving the increase in eccentricity in Fig.~\ref{fig:contour} and the transition between the two is shown.
First there is a region near the top of the plots ($i=180^\circ$)  where the apsidal precession driven by the companion adds to the GR precession causing the stationary inclination to be larger than $90^\circ$ and the largest eccentricity comes from librating orbits near retrograde coplanar.  The second region is near $i=90^\circ$ where the apsidal precession driven by the companion is larger than the GR precession and the stationary state has moved to less than $90^\circ$ and large eccentricity changes can be driven by near polar orbits.
For $m=1 \,\rm M_\odot$ at $r=10 \,a_{\rm b}$, the precession from the companion is about 19\% of that from GR  for $i=180^\circ$  while at $i=90^\circ$  it is about 57\% and in the retrograde direction. At higher mass the precession from the companion becomes more important.  So as the companion mass increases the region of high eccentricity near retrograde coplanar gets smaller and moves to lower radius while the region of high eccentricity driven by precession in the opposite direction caused by the companion near polar orbits gets larger and moves outward.

The lower left panel in Fig.~\ref{fig:contour} shows the effect of the companion eccentricity. We take the
$a_{\rm b}=50\,\rm R_\odot$ case with  $m=4 \,\rm M_\odot$  and vary the eccentricity for the companion with values $e = 0.2$, $0.4$, $0.6$ and $0.8$.  The results change very little, the most notable difference is that unstable companion orbits start
appearing when the periastron separation reaches about $3 \,a_{\rm b}$, which occurs for $r\lesssim 7.5 \,a_{\rm b}$ for $e=0.6$ and at $r\lesssim 15 \,a_{\rm b}$ for $e=0.8$.  These unstable radii are  consistent with the findings of \cite{Chen2020}.  For the stable three-body orbits at larger semi-major axis, the maximum eccentricity of the binary increases weakly with the eccentricity of the outer body. Therefore the maximum binary eccentricities shown in cases with a circular orbit binary may be increased with an eccentric orbit companion. 

Finally, the lower right panel in Fig.~\ref{fig:contour} explores the effect of an unequal mass binary with $m_1=30\,\rm M_\odot$ and $m_2=8\,\rm M_\odot$. The companion mass is scaled by the same factor as the total binary mass such that  $m=38/60\,\rm M_\odot$.  This is the same companion mass ratio as the upper right panel ($m=1\,\rm M_\odot$) but now with an asymmetric binary.  The unequal binary mass can drive a higher change in binary eccentricity. Therefore the results we have presented for the equal mass binary represent the  lower limit to the binary eccentricity growth.

The simulations we have considered all started with an initial binary eccentricity of $e_{\rm b}=0.2$. However, we have also examined selected cases with  eccentricities  down to $e_{\rm b}=0.01$ and up to $e_{\rm b}=0.6$ found they result in similarly high maximum  binary eccentricities. Additionally, these systems were all run with an initial longitude of ascending node $\Omega=90^\circ$, but we
have also run simulations with $\Omega=0$ and found very similar results. 

\section{Analytic Results}
\label{analytic}

We now find analytically the nodal precession rate for the three body system based on the quadrupole order expansion of the Hamiltonian.  In the frame of the binary, the nodal precession of the companion body is
\begin{equation}
    \dot \phi_c = \dot \phi_{c,\rm quad} - \dot \omega_{\rm GR}- \dot \varpi_{b,\rm quad},
    \label{main}
\end{equation}
where $\dot \phi_{c,\rm quad}$ is the nodal precession driven by the inner body, $\dot \omega_{\rm GR}$ is the apsidal precession of the binary driven by
GR and $\dot \varpi_{b,\rm quad}  = \dot \Omega_{2,\rm quad}+\dot \omega_{2,\rm quad}$ the precession rate for the longitude of the periastron.

In the limit of a small mass companion, the nodal precession of the outer body driven by the inner body is
\begin{equation}
    \dot \phi_{c,\rm quad}=-\frac{3}{4}\Omega_{\rm b}\frac{m_1m_2}{m_{\rm b}^2}\left(\frac{a_{\rm b}}{r}\right)^{7/2} F_1,
\end{equation}
where 
\begin{align}
\Omega_{\rm b}&=\sqrt{\frac{k^2 m_{\rm b}}{a_{\rm b}^3}}
\end{align}
and
\begin{align}
 F_1&=\cos(i) \frac{2+3 e_{\rm b}^2 + 5 e_{\rm b}^2 \cos(2\phi_2)}{2} 
\end{align}
\citep{Innanen1997,Childs2023}, where $k^2$ is the gravitational constant.

The apsidal precession of the binary driven by GR is 
\begin{equation}
    \dot \omega_{\rm GR}=3k^3\frac{(m_1+m_2)^{3/2}}{a_{\rm b}^{5/2}c^2(1-e_{\rm b}^2)}
\end{equation}
\citep[][]{Zanardi2018,lepp2022}.
The precession rate for the longitude of the periastron of the binary caused by the small mass companion is given by 
\begin{equation}
    \dot \varpi_{b,\rm quad}  = \frac{3}{4}\Omega_{\rm b}\frac{m}{(m_1+m_2)}\left(\frac{a_{\rm b}}{r}\right)^3 F_2,
\end{equation}
where  $F_2$ is given by
\begin{multline}
F_2=(1-e_{\rm b}^2)^{-1/2} \left[\right. 2-2e_{\rm b}^2+5(e_{\rm b}^2-\sin^2{i})\sin^2{\omega_{\rm b}} \\
 -\cos{i} (1-e_{\rm b}^2 + 5e_{\rm b}^2\sin^2{\omega_{\rm b}} )\left.\right]
\end{multline}
and $\omega_{\rm b}$ is the argument of periastron of the binary.  While in general this function is complicated because of the $\omega_{\rm b}$ part, it reduces to simple forms for coplanar or  polar companions.  
For coplanar, $i=0^\circ$  or  $180^\circ$, the $\omega_{\rm b}$ terms cancel while for a polar companion $i=90^\circ$, $\omega_{\rm b}=90^\circ$.
In these two cases
\[
F_2=
\begin{cases}
(1-e_{\rm b}^2)^{1/2}  &\text{ coplanar}\\
-3(1-e_{\rm b}^2)^{1/2}  &\text{ polar}.
\end{cases}
\]

The stationary inclination occurs where the nodal precession rate of the outer body is equal to the apsidal precession rate of the binary and therefore we solve $\dot \Omega_2=0$ with equation~(\ref{main}).
We can find the stationary inclination by setting $\dot\phi_2=0$ and $\phi_2=90^\circ$ giving
\begin{multline}
    \cos(i_s) =
    -(\dot\Omega_{\rm GR}+\dot{\varpi}_{b,\rm quad} )\times\\
       \frac{4}{3k} \frac{(m_1+m_2)^{3/2}}{m_1m_2} \frac{{r}^{7/2}}{a_{AB}^2}
  \frac{1}{(1+4e_{\rm b}^2)}\,.
    \label{eq:is}
\end{multline}
If  $\dot{\varpi}_{b,\rm quad}=0$ then this reduces to equation~(4) in \cite{lepp2022} or if $\dot\Omega_{\rm GR}=0$ then it reduces to equation~(8) in \cite{Childs2023}.

\subsection[Massive third body: i=180]{Stationary inclination of $i_{\rm s}=180^\circ$ }

We can calculate a critical radius where the stationary inclination is equal to $i_{\rm s}=180^\circ$. For larger radius, $r>r_{\rm c}$, there is no librating region. 
The critical radius including the apsidal precession driven by GR and by the mass of the third body is calculated with
\begin{equation}
     \frac{r_{\rm c1}}{a_{\rm b}} = \left[\frac{ \big(\frac{3\Omega_{\rm b}}{4}\big) 
    \big(\frac{m_1m_2}{m_{\rm b}^2}\big) (1+4e_{\rm b}^2)}{-\dot\omega_{\rm GR}-\dot\varpi_{b, \rm quad}}\right]^{2/7}.
\end{equation}
This critical radius depends on the orbital radius of the companion and so must be solved self consistently.  These are calculated for an inclination of 180$^\circ$ where we may use the simpler expression for $F_2$ for the coplanar case.  This radius is 
marked by the right most black arrows on the $i=180^\circ$ inclination in Fig.~\ref{fig:contour}. This is the largest radius for a librating orbit, although this is much larger than the orbits which give the largest eccentricity changes.  

\subsection[Librating orbits reach i=180]{Librating orbit at $i=180^\circ$ }

In order to estimate where the maximum eccentricity occurs,  we instead find the critical radius at which the librating region just reaches $i=180^\circ$. This is calculated by finding where $i=\phi_2=180^\circ$, $\phi_c=0$ and
$\dot\phi_c=0$ to give  
\begin{equation}
     \frac{r_{\rm c2}}{a_{\rm b}} = \left[\frac{ \big(\frac{3\Omega_{\rm b}}{4}\big) 
    \big(\frac{m_1m_2}{m_{\rm b}^2}\big) (1-e_{\rm b}^2)}{-\dot\omega_{\rm GR}-\dot\varpi_{b, \rm quad}}\right]^{2/7}.
\end{equation}
Again the precession caused by the companion changes with radius and so this must be solved self-consistently.   The radius is shown by the left most arrows on the $180^\circ$ inclinations in Fig.~\ref{fig:contour}. These are seen to be fairly close to the peak in eccentricity growth for a retrograde orbit.

\subsection[Massive third body: i=90]{Stationary inclination of $i_{\rm s}=90^\circ$ }
\label{i90}

We can also calculate a critical radius for the polar ($i=90^\circ$) case, $r_{\rm c3}$.  In this case, the precession caused by the companion is in the opposite direction to that caused by GR. The radius where the two precessions cancel is a stationary state and $i=i_{\rm s}=90^\circ$. This is marked by a black arrow on the $i=90^\circ$ inclination in Fig.~\ref{fig:contour}. This critical radius shows a good correlation with the maximum radius at which eccentricity growth occurs for a polar companion. At smaller radii the precession from the companion dominates and  $i_{\rm s}<90^\circ$.  Here, orbits which start with  $i=90^\circ$ librate about the stationary state and cause an increase in eccentricity of the binary.  At larger radii, the GR precession dominates and $i_{\rm s}>90$. Orbits which start with  $i=90^\circ$  orbit about the stationary state and drive the eccentricity of the binary to lower values.

\subsection{Scaling with binary mass and semi-major axis}
\label{scaling}

The analytic radii calculated in the previous subsections all share a common scaling law. If we take all the masses $m_1,m_2,m\propto m_{\rm b}$ and the orbital radius of the companion to be $r\propto a_{\rm b}$ then we can investigate how the system scales in relation to these two parameters. We ignore the binary eccentricity dependence since the critical radii are relatively insensitive to this.  In both $r_{c1}/a_{\rm b}$ and $r_{c2}/a_{\rm b}$ the numerator scales as $\sqrt{m_{\rm b}/a_{\rm b}^3}$, $\dot\omega_{GR}$ scales as $\sqrt{m_{\rm b}^3/a_{\rm b}^5}$ and $\dot\varpi_{b, \rm quad}$ scales as $\sqrt{m_{\rm b}/a_{\rm b}^3}$.  If we now further impose the same scaling on mass as on length, so that $m_{\rm b}\propto a_{\rm b}$, 
the critical radii as a ratio to $a_{\rm b}$ are constant.   The critical radius $r_{\rm c3}$ (Section~\ref{i90})  scales in the same manner.

The entire system scales as long as both the masses and lengths are changed by the same factor.  For example, if we multiply everything by a factor of two, then our standard parameters become two $m_1=m_2=60\,\rm M_\odot$ black holes with $a_{\rm b}=20$, $60$, $100$ and $500 \,\rm R_\odot$ and companions with mass 1, 2, 4 and $8\,\rm M_\odot$. With these parameters, Fig~\ref{fig:contour} and Fig~\ref{fig:contour2} are unchanged.  Though this scaling is only strictly true for our analytical results, numerical tests suggest it works for the whole system. 
This suggests that our results are widely applicable to systems of any binary mass, provided that the binary semi-major axis and tertiary mass are appropriately scaled.

\section{Merger times}
\label{merger}

\begin{figure*}
\begin{center}
\includegraphics[width=.9\columnwidth]{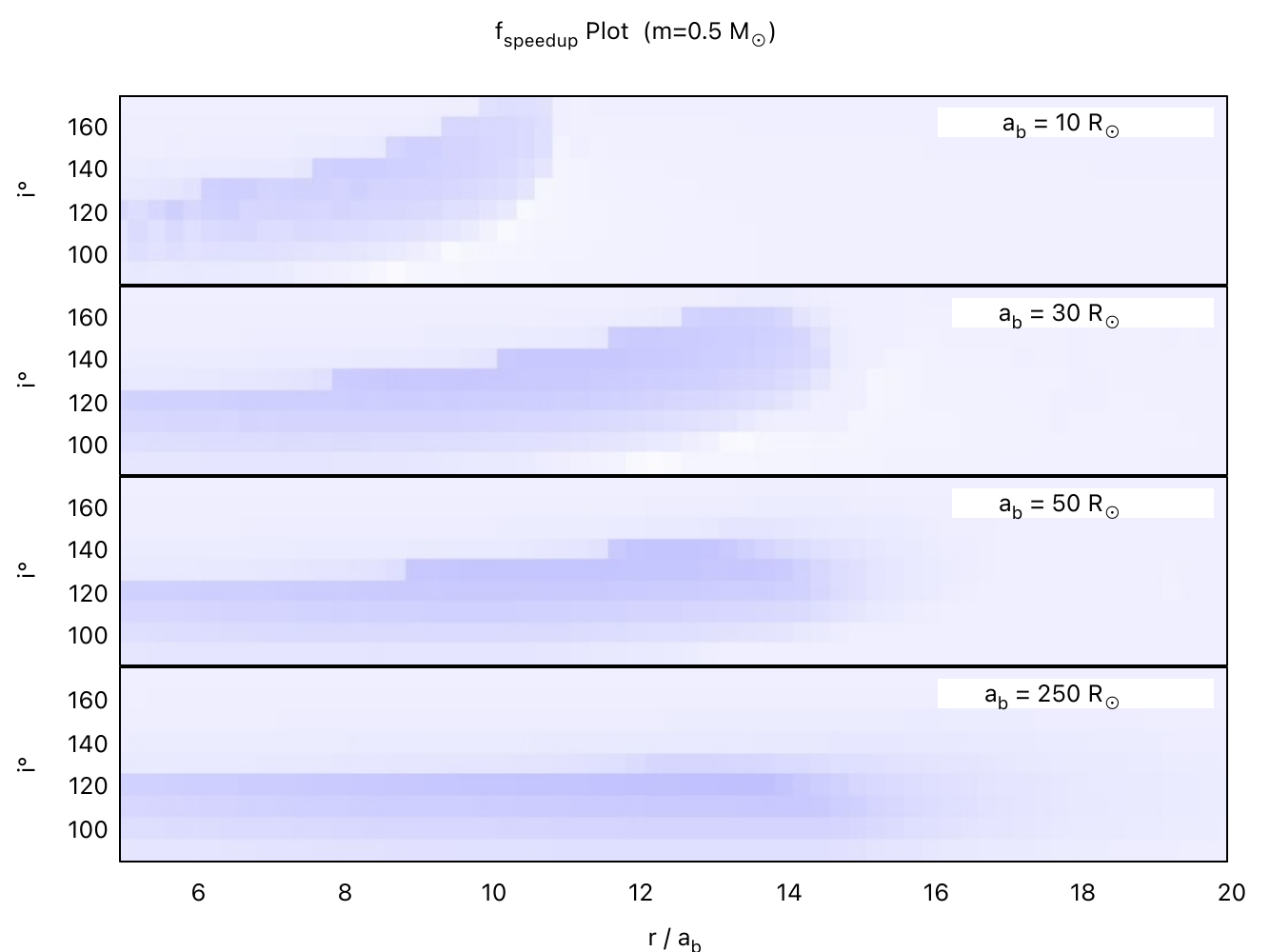}
\includegraphics[width=.9\columnwidth]{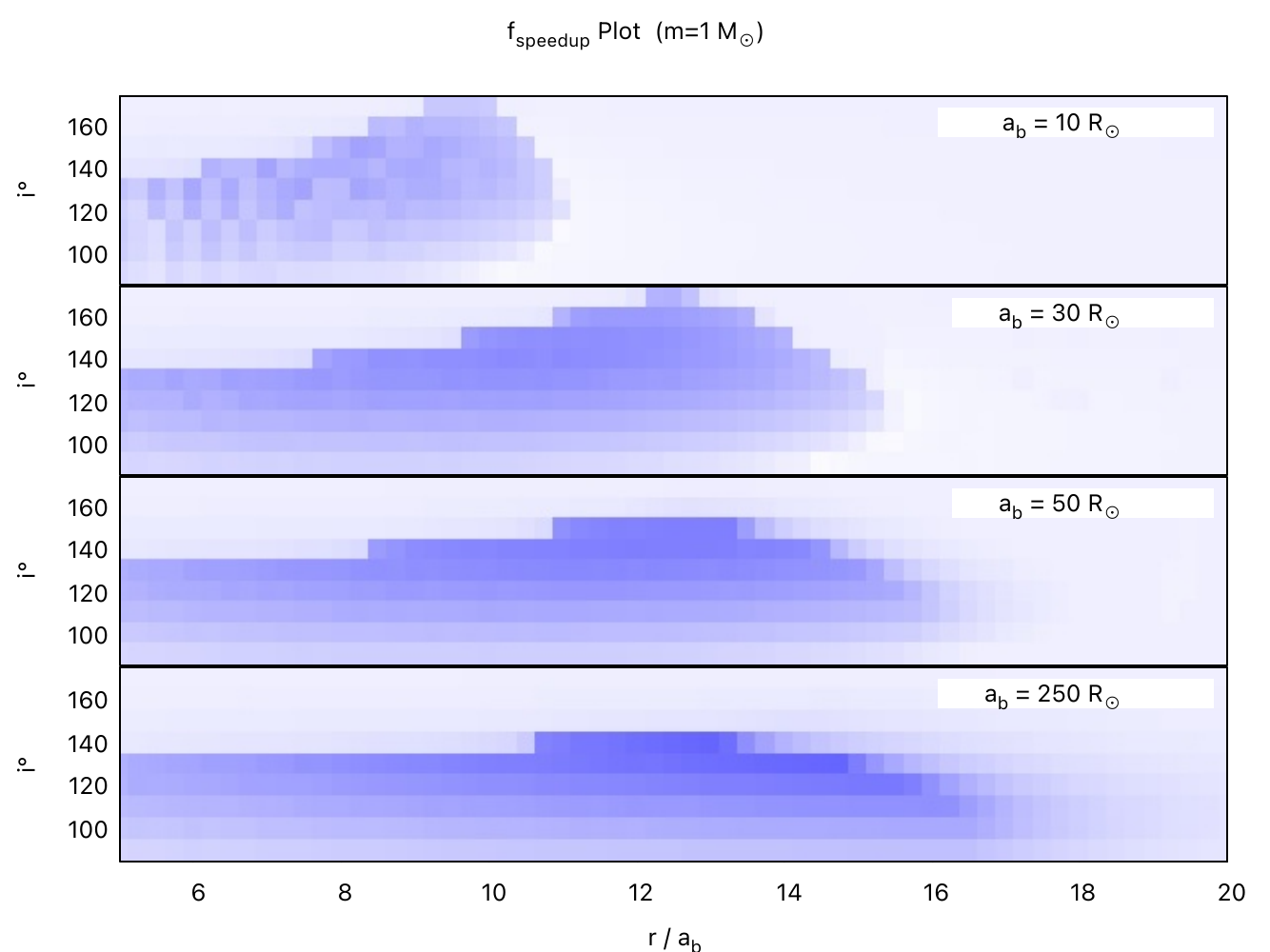}
\includegraphics[height=.6\columnwidth]{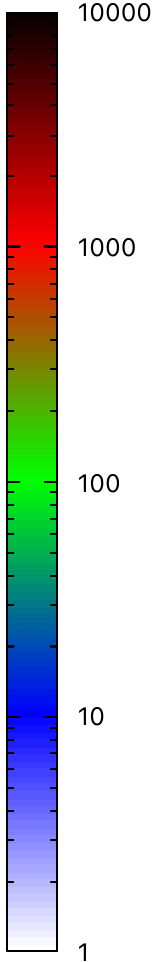} \raisebox{1.0in}{$f_{speedup}$}
\includegraphics[width=.9\columnwidth]{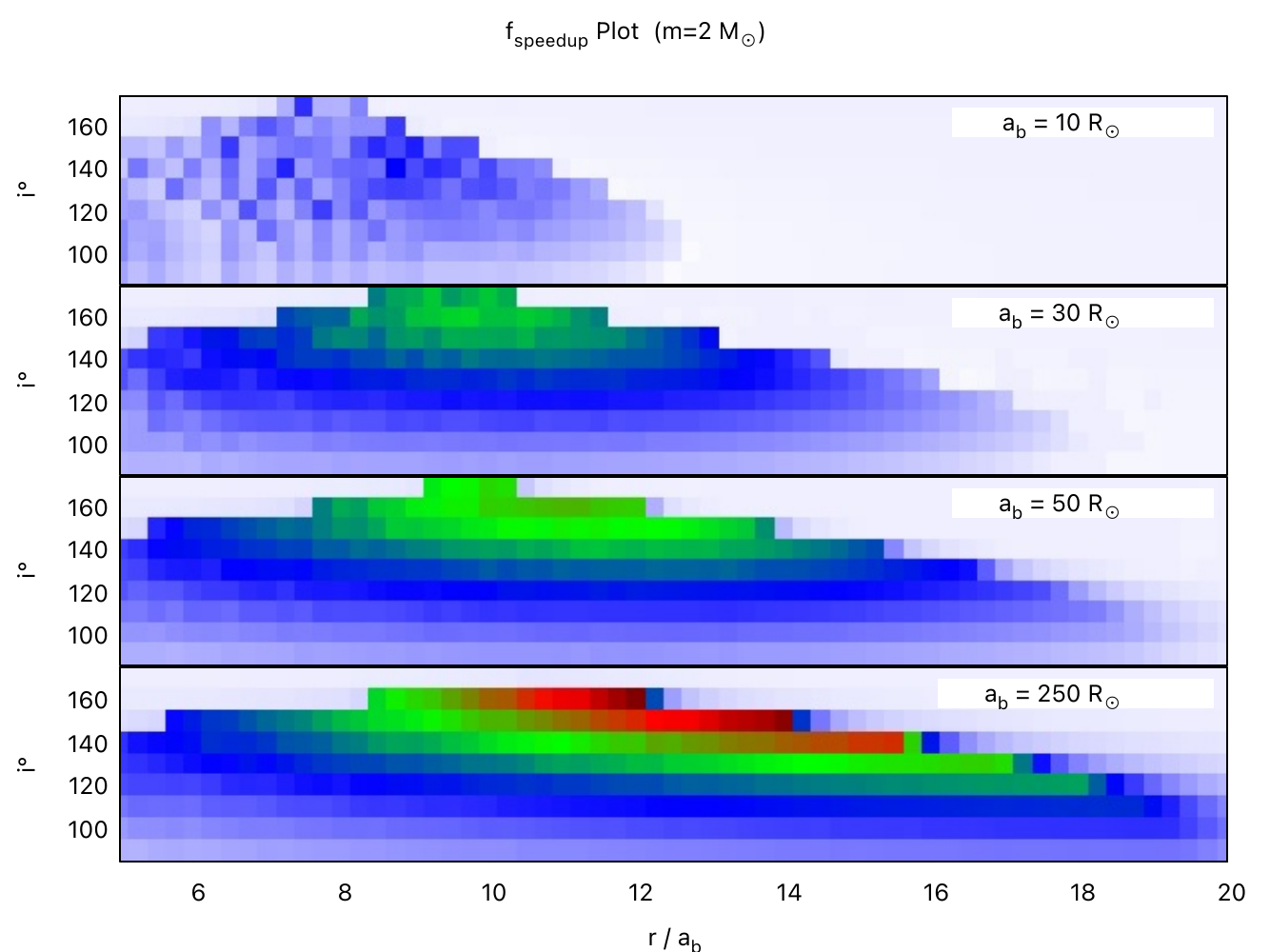}
\includegraphics[width=0.9\columnwidth]{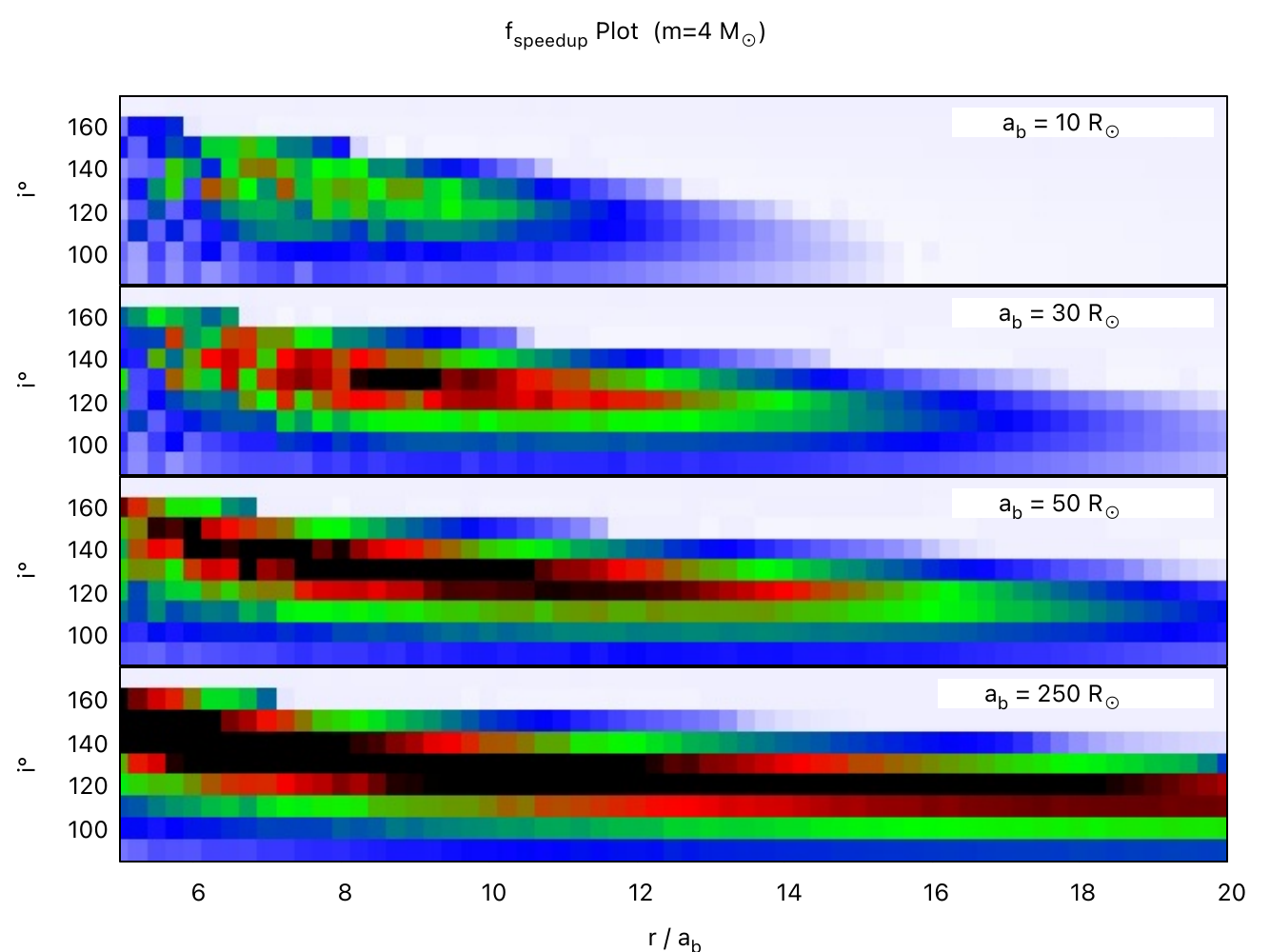}
\includegraphics[height=.6\columnwidth]{key.pdf} \raisebox{1in}{$f_{speedup}$}
\includegraphics[width=.9\columnwidth]{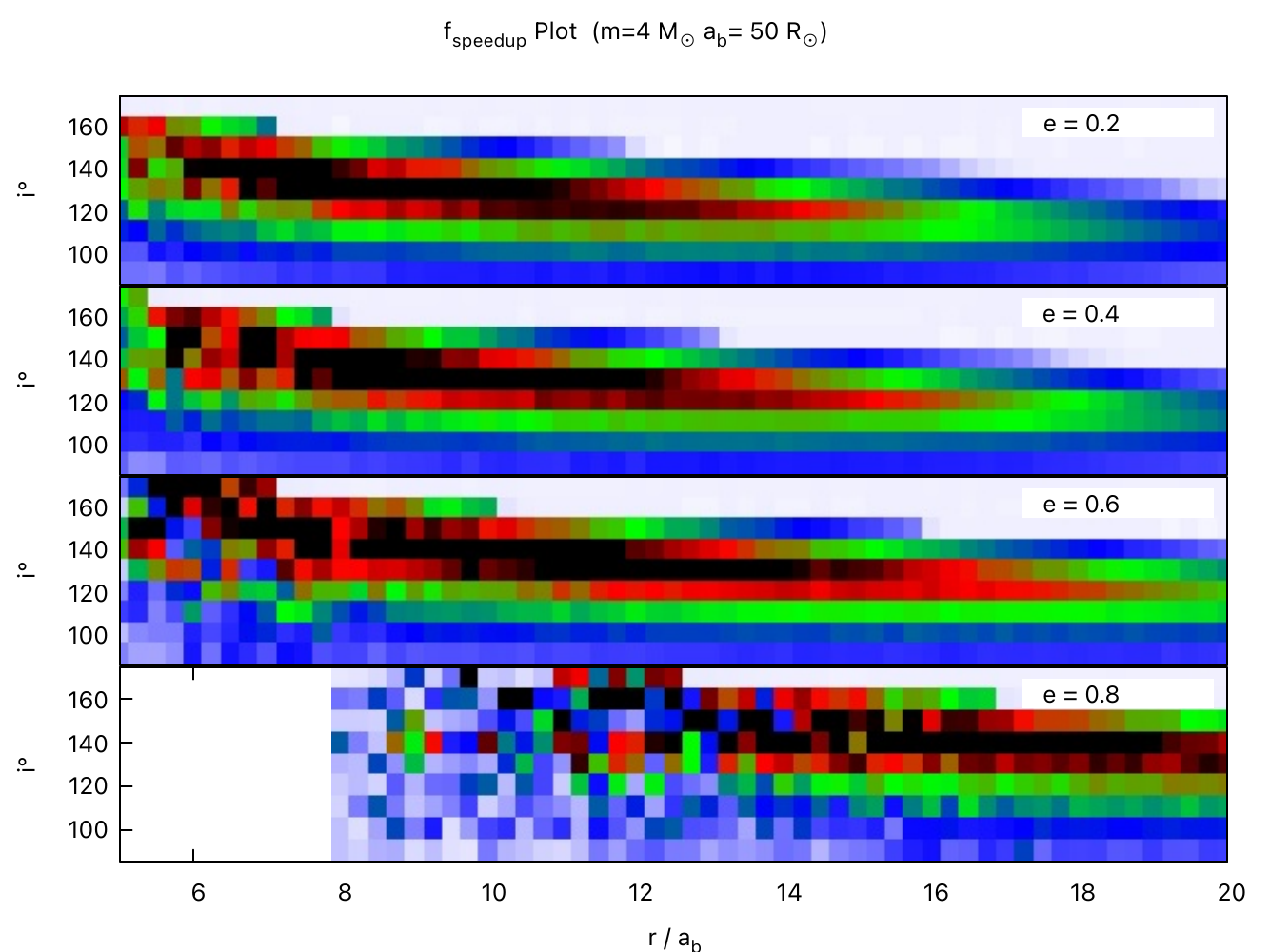}
\includegraphics[width=0.9\columnwidth]{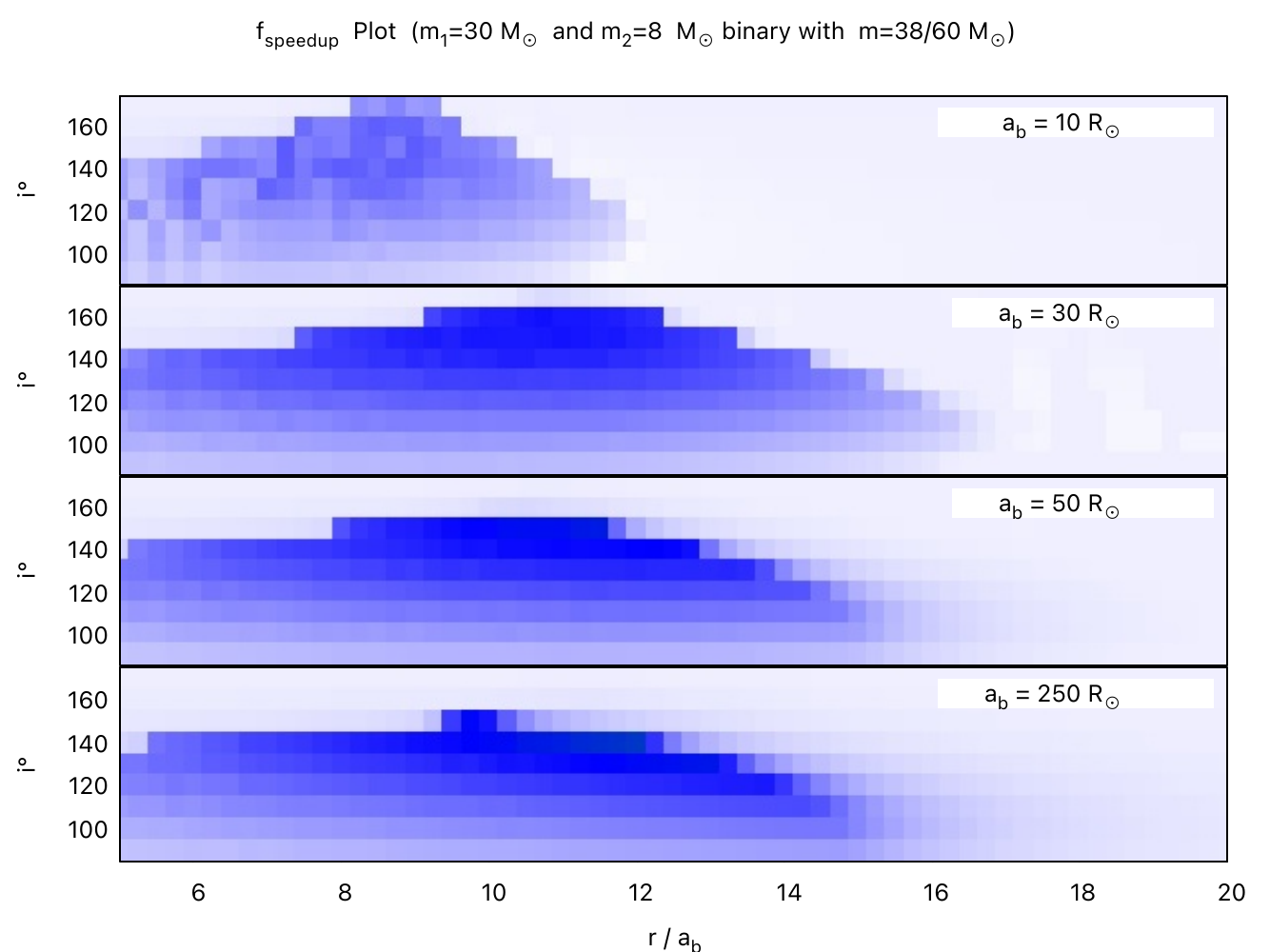}
\includegraphics[height=.6\columnwidth]{key.pdf} \raisebox{1in}{$f_{speedup}$}
\end{center}
\caption{Same as Fig.~\ref{fig:contour} except the colors represent  the value of $f_{\rm speedup}$ given by equation~(\ref{1/F}).  This shows how many times faster the binary merger is expected to be due to the KL cycle compared to a circular orbit binary.  The color black is used for all speed-up factors greater than 10,000 though individual pixels may have significantly faster speed-ups.  The largest speed-up in the $m=4\,\rm M_\odot$ panel for $a_{\rm b}=30$, $50$ and  $250 \,\rm R_\odot$ are $7\times10^4$, $8\times10^5$ and  $7\times10^8$, respectively. }
\label{fig:contour2}
\end{figure*}

For a binary that is not undergoing KL oscillations, the evolution of the semimajor axis $a_{\rm b}$ and of the eccentricity $e_{\rm b}$ due to gravitational wave radiation can be written as 
\begin{eqnarray}
 \frac{da_{\rm b}}{dt} & = & -\frac{64}{5} \frac{k^6 \mu m_{\rm b}^2}{c^5 a_{\rm b}^3} \frac{1}{(1-e_{\rm b}^2)^{7/2}} \left(1+\frac{73}{24}e_{\rm b}^2 + \frac{37}{96} e_{\rm b}^4\right), \label{eq:da/dt} \\
 \frac{de_{\rm b}}{dt} & = & -\frac{304}{15} \frac{k^6 \mu m_{\rm b}^2}{c^5 a_{\rm b}^4} \frac{e_{\rm b}}{(1-e_{\rm b}^2)^{5/2}} \left(1+\frac{121}{304}e_{\rm b}^2\right) \label{eq:de/dt}
\end{eqnarray}
\citep{peters64,maggiore18},
where $m_{\rm b}=m_1+m_2$ is the total mass and $\mu = m_1m_2/m_{\rm b}$ is the reduced mass.  Integrating the orbit from initial values $a_{\rm b}(t_0)=a_0$ and $e_{\rm b}(t_0)=e_0$ until merger with $a_{\rm b}(t) = 0$ or $e_{\rm b}(t)=0$ (circular orbit at the merger), one can obtain the merger time $\tau=t-t_0$ as a function of $a_0$ and $e_0$ as
\begin{equation}
    \tau(a_0,e_0)=\frac{15}{304} \frac{c^5}{k^6 m_{\rm b}^2\mu} \int_0^{e_0} de_{\rm b} \frac{a_{\rm b}^4(e_{\rm b}) (1-e_{\rm b}^2)^{5/2}} {e_{\rm b}\left(1+\frac{121}{304} e_{\rm b}^2\right)},
\label{eq:tau}
\end{equation}
where $a_{\rm b}$ as a function of $e_{\rm b}$ can be integrated analytically based on Eqs. (\ref{eq:da/dt}) and (\ref{eq:de/dt}) as
\begin{equation}
    a_{\rm b}(e_{\rm b}) = c_0 g(e_{\rm b}), 
    %c_0 \frac{e^{12/19}} {1-e^2} \left(1+\frac{121}{304}e^2 \right)^{870/2299},
\label{eq:a(e)}
\end{equation}
where 
\begin{eqnarray}
 g(e_{\rm b}) & = & \frac{e_{\rm b}^{12/19}}{1-e_{\rm b}^2} \left(1+\frac{121}{304} e_{\rm b}^2 \right)^{870/2299}, 
\end{eqnarray}
$c_0$ depends on the initial condition $a_{\rm b}=a_0$ and when $e_{\rm b}=e_0$. Expressing $\tau$ in terms of orbital period $P_{\rm b}$, the merger time can be expressed as
\begin{eqnarray}
    \tau(a_0,e_0) & \simeq & (9.829 \ {\rm Myr}) \left(\frac{P_{\rm b,0}}{1 \rm {\rm hr}}\right)^{8/3}\left(\frac{\rm M_\odot}{m_{\rm b}}\right)^{2/3} \nonumber \\
    &\times & \left(\frac{\rm M_\odot}{\mu}\right) F(e_0),
\end{eqnarray}
where 
\begin{eqnarray}
 F(e_0) & = & \frac{48}{19}\frac{1}{g^4(e_0)} \int_0^{e_0} de_{\rm b} \frac{g^4(e_{\rm b}) (1-e_{\rm b}^2)^{5/2}} {e_{\rm b} (1+\frac{121}{304} e_{\rm b}^2)}.
% g(e) & = & \frac{e^{12/19}}{1-e^2} \left(1+\frac{121}{304} e^2 \right)^{870/2299}. 
\end{eqnarray}
The merger time speed-up of an orbit compared to a circular orbit binary can be calculated with $1/F$. For example,  $F(0.6) \simeq 0.206$, corresponding to a speed-up factor of $f_{\rm speedup}\simeq 4.86$ while $F(0.95)=0.000364$ which corresponds to speed-up factor of $f_{\rm speedup}\simeq 2750$.

Since the KL oscillation timescale is much shorter than the binary orbital period, the speed-up  during the KL oscillation is well approximated by the average speed-up factor given by
\begin{equation}
    f_{\rm speedup}=\left<\frac{1}{F}\right>,
    \label{1/F}
\end{equation}
where $<>$ denotes the average value over a KL oscillation period.
This is a good approximation because the  speed-up is a measure of the rate of gravitational wave emission and the 
binary orbital period is much shorter than the KL period.  As a result, over a binary orbital period, the energy loss to gravitational waves is approximately the same as that of a system not undergoing KL oscillations.  By averaging over a KL cycle we are averaging the energy loss.

Fig.~\ref{fig:contour2} shows the same simulations as Fig.~\ref{fig:contour} but the pixels are coloured by $f_{\rm speedup}$ given by equation~(\ref{1/F}).  For third body masses of $m=0.5\,\rm M_\odot$ and $m=1\,M_\odot$, the eccentricities cycle to high values but do not spend enough time there for a significant average speed up.  For our parameter space this leaves  speed-up factors $\lesssim 10$.  The reason for this can be seen in Fig.~\ref{fig:ecc} where one can see that although it cycles above 0.6 where the $1/F$ gets above 5 it only does so for brief portion of the cycle and so the average speed-up remains low.

For higher masses of the third body, the KL cycle gets to even higher eccentricities. 
The fastest speed-up for both 2 and 4 M$_\odot$ occur at mid inclinations. Even though the highest eccentricities occur for inclinations of near 180$^\circ$, the highest speed-ups occur at smaller inclinations which have faster KL cycles and spend more time near their highest inclinations, as can be seen in Fig.~\ref{fig:ecc}.  For $m=2\,\rm M_\odot$ we find speed-ups of order 100 and for $m=4\,\rm M_\odot$ there are speed-ups on the order of over 10,000.

In the quadrupole approximation, \cite{Liu2017} find that the merger rate is sped up by a factor 
\begin{equation}
    f_{\rm speedup}\approx (1-e_{\rm max}^2)^{-\alpha}
    \label{ll}
\end{equation}
compared to a circular orbit binary. The value of $\alpha$ depends on $e_{\rm max}$. For $e_{\rm max}$ in the range $(0.6,0.8)$, $\alpha=2$, for $e_{\rm max}$ in the range $(0.8,0.95)$, $\alpha=2.5$ and for $e_{\rm max}>0.95$, $\alpha=3$,  For $e_{\rm max}=0.6$ this gives $f_{\rm speedup}=2.4$ and for $e_{\rm max}=0.95$ this gives $f_{\rm speedup}\approx 1000$.  For comparison, we consider an example of our $m=2\,\rm  M_\odot$ and $a_{\rm b}=50 \,\rm R_\odot$ panel for conditions that give $e_{\rm max}\approx 0.6$.  Such orbits have slightly slower speed-ups between $1.69$ and $2.23$, with the slowest coming from $i\approx 130^\circ$ and the fastest coming from $i\approx 90^\circ$. 
%For conditions giving  $e_{\rm max}\approx 0.95$ we have significantly lower speed-ups of only $60$ to $120$. However, these come from inclinations of at least $i=150^\circ$ where the KL oscillation times are long and very little time is spent at $e_{\rm max}$.  
We have made Fig.~\ref{fig:contour2} using equation~(\ref{ll}) and find very similar results. We find that Equation~(\ref{1/F}) gives speed-up factors that are less than or approximately equal to those calculated by equation~(\ref{ll}).   Therefore  equation~(\ref{ll}) leads to slightly larger regions of high speed-up.

\section{Conclusions}
\label{concs}

A companion star orbiting around a black hole binary can drive eccentricity growth of the black hole binary through KL oscillations.   With $n$-body simulations and analytic methods including the effects of GR we have shown that a companion mass of just a few percent of the binary mass can induce large KL eccentricity oscillations if the orbit is closer to retrograde than to prograde.  The large increase in eccentricity can reduce the merger timescale by orders of magnitude compared to a circular orbit binary.  
As the companion mass increases,  the region of large merger speed-up factors from induced high eccentricities moves from nearly retrograde toward more polar orbits. For companion masses greater than about 20\% of the binary mass the mechanism is more effective for polar orbits as has been seen before \citep{Liu2018,Liu2019,Liu2019b}.
Because low mass objects are much more numerous that high mass objects,  this mechanism may prove important in the evolution of black hole binary systems.

\section*{acknowledgements}
We thank an anonymous referee for useful comments. We acknowledge support from NASA through grant numbers 80NSSC21K0395 and 80NSSC23M0104.

\bibliographystyle{aasjournal}
\bibliography{bh} % if your bibtex file is called example.bib

\end{document}